\documentclass[useAMS,usenatbib]{mn2e}
\usepackage{times}

\title[Radio imaging of MM~18423+5938]{A new perspective on the submillimetre galaxy MM\,18423+5938 at redshift 3.9296 from radio continuum imaging}
\author[J. P. McKean et al.]{J. P. McKean,$^1$\thanks{mckean@astron.nl} A. Berciano Alba,$^1$ F. Volino,$^2$ V. Tudose,$^{1,3,4}$ M. A. Garrett,$^{1,5,6}$ 
\newauthor A. F. Loenen,$^{5}$ Z. Paragi$^{7,8}$ and O. Wucknitz$^2$\\
$^1$ASTRON, Oude Hoogeveensedijk 4, 7991 PD Dwingeloo, the Netherlands\\
$^2$Argelander-Institut f\"{u}r Astronomie, Auf dem H\"{u}gel 71, D-53121 Bonn, Germany\\
$^3$Astronomical Institute of the Romanian Academy, Cutitul de Argint 5, RO-040557 Bucharest, Romania\\
$^4$Research Center for Atomic Physics and Astrophysics, Atomistilor 405, RO-077125 Bucharest, Romania\\
$^5$Leiden Observatory, Leiden University, Postbus 9513, 2300 RA Leiden, the Netherlands\\
$^6$Centre for Astrophysics and Supercomputing, Swinburne University of Technology, Hawthorn, Victoria 3122, Australia\\
$^7$Joint Institute for VLBI in Europe (JIVE), Postbus 2, 7990 AA Dwingeloo, the Netherlands\\
$^8$MTA Research Group for Physical Geodesy and Geodynamics, P.O. Box 91, H-1521 Budapest, Hungary}
\begin{document}

\date{Accepted 2011 March 2. Received 2011 February 28; in original form 2011 January 26}

\pagerange{\pageref{firstpage}--\pageref{lastpage}} \pubyear{2011}

\maketitle

\label{firstpage}

\begin{abstract}
The bright sub-mm galaxy MM\,18423+5938 at redshift 3.9296 has been predicted from mid-infrared and millimetre photometry to have an exceptionally large total infrared (IR) luminosity. We present new radio imaging at 1.4 GHz with the Westerbork Synthesis Radio Telescope that is used to determine a radio-derived total IR luminosity for MM\,18423+5938 via the well established radio--far-infrared correlation. The flux density is found to be $S_{\rm 1.4~GHz}=$~217\,$\pm$\,37~$\mu$Jy, which corresponds to a rest-frame luminosity density of $L_{\rm 1.4~GHz}=$~2.32\,$\pm$\,0.40\,$\times$\,10$^{25}$~$\mu^{-1}$~W~Hz$^{-1}$, where $\mu$ is the magnification from a probable gravitational lens. The radio-derived total IR luminosity and star-formation rate are $L_{8-1000~\mu m}$~=~5.6$^{+4.1}_{-2.4}$\,$\times$\,10$^{13}$~$\mu^{-1}$~L$_{\odot}$ and SFR~=~9.4$^{+7.4}_{-4.9}$\,$\times$\,10$^{3}$~$\mu^{-1}$~M$_{\odot}$~yr$^{-1}$, respectively, which are $\sim$9 times smaller than those previously reported. These differences are attributed to the IR spectral energy distribution of MM\,18423+5938 being poorly constrained by the limited number of reliable photometric data that are currently available, and from a previous misidentification of the object at 70~$\mu$m. Using the radio derived total IR luminosity as a constraint, the temperature of the cold dust component is found to be $T_{d}\sim$~24$^{+7}_{-5}$~K for a dust emissivity of $\beta=$~1.5\,$\pm$\,0.5. The radio-derived properties of this galaxy are still large given the low excitation temperature implied by the CO emission lines and the temperature of the cold dust. Therefore, we conclude that MM\,18423+5938 is probably gravitationally lensed.
\end{abstract}


\section{Introduction}

Sub-millimetre (sub-mm) selected galaxies have become a well established population of heavily dust obscured objects at high redshift (e.g. \citealt*{blain02}). They are characterized by their extremely large total infrared (IR) luminosities ($L_{\rm8-1000~\mu m}\sim$~2\,$\times$\,10$^{12}$~L$_{\odot}$;  e.g. \citealt{aretxaga07}) and very large star-formation rates (SFR~$\sim$~400~M$_{\odot}$~yr$^{-1}$; e.g. \citealt{coppin08}). Sub-mm galaxies are important for galaxy formation studies because they are thought to be the progenitors of present day ellipticals (e.g. \citealt{swinbank06}) and because they are believed to make up a significant fraction of the star-formation density at high redshift (e.g. \citealt{wardlow10}). Although the bulk of the radio-selected sub-mm galaxy population is located at $\bar z \sim$~2.5 \citep{chapman05}, there is growing evidence from surveys carried out at mm-wavelengths that there is also a significant population of sub-mm galaxies at higher redshifts \citep{bertoldi07,younger09}. However, only seven $z>$~3 sub-mm galaxies have so far been confirmed spectroscopically  \citep{capak08,coppin09,daddi09a,daddi09b,knudsen10,lestrade10}.

The bright sub-mm galaxy MM\,18423+5938 was recently discovered by \citet{lestrade10} with MAMBO2 (Max-Planck Millimetre Bolometer 2).  This rare object has 1.2, 2 and 3 mm flux-densities of 30, 9 and 2 mJy, respectively. Follow-up spectroscopic observations found the redshift to be $z =$~3.9296 from CO (4--3, 6--5 and 7--6 transitions) and C\,I emission lines. The spectral energy distribution of the CO lines is consistent with a moderate starburst, and excludes heating by a dominant active galactic nucleus (AGN). MM\,18423+5938 was not detected by {\it IRAS} at 60 and 100~$\mu$m, or with {\it Spitzer} at 24~$\mu$m. However, a source lying 3$\sigma$ from the MAMBO2 position was found in {\it Spitzer} imaging at 70~$\mu$m with a flux density of 31 mJy. From a fit to the observed spectral energy distribution, using 4 detections and 3 upper-limits at mm and mid-infrared (MIR) wavelengths, the total IR luminosity of MM\,18423+5938 was estimated to be $L_{\rm 8-1000~\mu m}=$~4.8\,$\times$\,10$^{14}$~$\mu^{-1}$~L$_{\odot}$ \citep{lestrade10}. Given the extremely large total IR luminosity, which is an order of magnitude higher than for rare hyper-luminous sub-mm galaxies, it is expected that there is a foreground gravitational lens magnifying MM\,18423+5938 (the factor $\mu$ above). The large observed flux density and the possible large lensing magnification of MM\,18423+5938 provides a great opportunity to investigate the properties of a very high redshift mm-selected galaxy in detail.

In this letter, sensitive radio continuum observations are used to investigate the nature of MM\,18423+5938. The main aim of these observations is to use the well established radio--far-infrared (FIR) correlation to independently determine the total IR luminosity of MM\,18423+5938 and hence, calculate a radio-derived star-formation rate. Throughout, we adopt an $\Omega_{\rm M} =$~0.3 and $\Omega_{\Lambda}=$~0.7 spatially flat Universe, with a Hubble constant of $H_{0} =$~70~km\,s$^{-1}$~Mpc$^{-1}$, and a solar luminosity of $L_{\odot}=$~3.939~$\times$~10$^{26}$ W.

\section{Observations \& Results}
\label{obs}

MM\,18423+5938 was observed with the Westerbork Synthesis Radio Telescope (WSRT) at 1.381 GHz on 2010 September 15, 16 and 17 for a total of 24 h. The observations were performed in the standard continuum mode, which consists of 8 spectral sub-bands, each with 20 MHz bandwidth,  64 spectral channels, and 2 polarizations. 3C\,286 and 3C\,48 were used to correct for the bandpass in each sub-band and to apply the flux-density calibration to the data. Radio frequency interference was first flagged and then the target data were self-calibrated to determine the amplitude and phase gain variations in the standard way using the Common Astronomy Software Applications ({\sc casa}) package. The imaging was carried out using uniform weighting, which gave a beam-size of 14.4 $\times$ 9.9 arcsec$^{2}$ at a position angle of $-$7.6 deg and a map rms of 21~$\mu$Jy~beam$^{-1}$. The expected map noise for an observation with the WSRT of 24 h duration is 8.4~$\mu$Jy~beam$^{-1}$. However, the image of MM\,18423+5938 is dynamic range limited due to the presence of a bright 137~mJy off-axis radio source that is about 10 arcmin south-east from the phase centre.

\begin{figure}
\begin{center}
\setlength{\unitlength}{1cm}
\begin{picture}(6,9.5)
\put(-1.3,0){\includegraphics{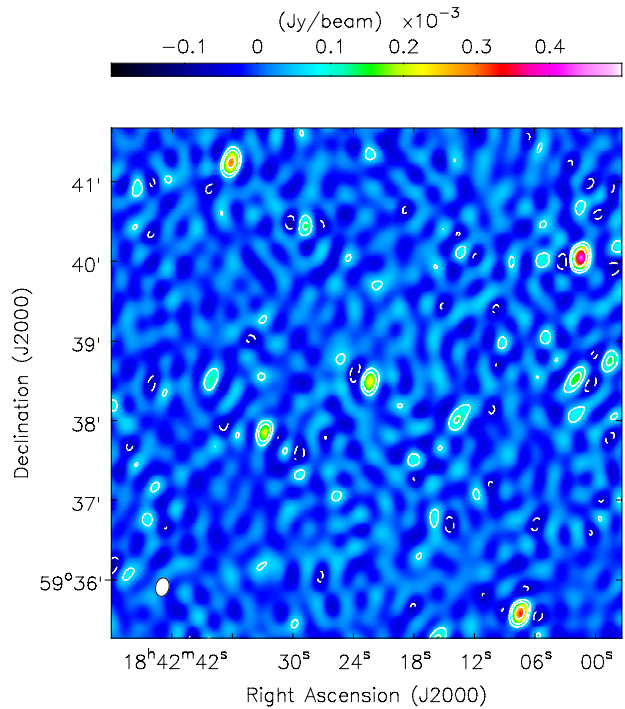}}
\end{picture}
\caption{The WSRT image of MM\,18423+5938 at 1.381 GHz. The galaxy is detected at the 11$\sigma$ level and is seen at the expected position in the centre of the map. The contours are ($-$3, 3, 6, 12, 24) $\times$ 21~$\mu$Jy~beam$^{-1}$, the rms map noise. The beam-size, shown by the white ellipse in the bottom left-hand corner, is 14.4 $\times$ 9.9 arcsec$^{2}$ at a position angle of $-$7.6 deg east of north.} 
\label{wsrt-image}
\end{center}
\end{figure}

The 1.381 GHz image of the region around MM\,18423+5938 is presented in Fig.~\ref{wsrt-image}. MM\,18423+5938 is detected at the 11$\sigma$ level and is found to have a peak surface brightness of $I_{\rm 1.4~GHz}=$~232\,$\pm$\,21~$\mu$Jy~beam$^{-1}$ at the position $\alpha_{\rm J2000}=$~18$^h$42$^m$22.32$^s$\,$\pm$\,0.06$^s$ $\delta_{\rm J2000}=$~$+$59\degr38\arcmin29.6\arcsec\,$\pm$\,0.6\arcsec. This radio position is in good agreement with and has better precision than the position measured at mm-wavelengths \citep{lestrade10}. A Gaussian fit to the radio emission in the image plane finds a 1.381-GHz flux density of $S_{\rm 1.4~GHz}=$~217\,$\pm$\,37~$\mu$Jy from MM\,18423+5938. The flux density that was found on each of the separate observing days is consistent with this measurement from the combined dataset. The de-convolved  full width at half maximum of the fitted Gaussian is $r_{\rm FWHM}=$~2.6$_{-2.6}^{+4.5}$~arcsec (c.f. with the beam-size above). Therefore, the radio emission detected from MM\,18423+5938 is unresolved on the scales probed here with the WSRT.

The rest-frame 1.4 GHz radio luminosity density of MM\,18423+5938 has been estimated using,
\begin{equation}
L_{\rm 1.4~GHz} = 4\pi {D_{L}}^2S_{\rm 1.4~GHz}(1+z)^{-{(1+\alpha)}}
\end{equation}
where $D_{L}$ is the luminosity distance, $S_{\rm 1.4~GHz}$ is the flux density, $z$ is the redshift and $\alpha$ is the two-point spectral index ($S_{\nu} \propto \nu^{\alpha}$). The spectral index is needed to apply the $k$-correction to the flux density and is taken to be $\alpha = -$0.8. This is the spectral index that is assumed for the radio--FIR correlation we use later and is consistent with the spectral index of a typical sub-mm galaxy at redshift $\sim$2 ($-$0.75\,$\pm$\,0.06; \citealt{ibar10}). The rest-frame 1.4 GHz radio luminosity density is found to be $L_{\rm1.4~GHz}=$~2.32\,$\pm$\,0.40\,$\times$\,10$^{25}$~$\mu^{-1}$~W~Hz$^{-1}$.  However, it should be noted that there is tentative evidence for the radio spectral indices of sub-mm galaxies steepening with redshift \citep{bourne10}, which would result in a higher luminosity density than has been calculated here; a change of $\pm$\,0.2 in the spectral index would result in a change of $\sim$35 per cent in the luminosity density.

The properties of MM\,18423+5938 from the WSRT observations and from our analysis presented below are summarized in Table \ref{table}.

\begin{table}
\begin{center}
\caption{The properties of MM\,18423+5938 from the WSRT observations and from a fit to a modified blackbody spectrum. Some parameters are expressed in terms of the unknown gravitational lens magnification $\mu$.}
\begin{tabular}{lll}
\hline
Right Ascension				& 18 42 22.32\,$\pm$\,0.06 		& $^h$ $^m$ $^s$\\		
Declination 					& $+$59 38 29.6\,$\pm$\,0.6 		& \degr~\arcmin~\arcsec\\
$I_{\rm 1.4~GHz}	$ 			& 232\,$\pm$\,21 				& $\mu$Jy~beam$^{-1}$\\
$S_{\rm 1.4~GHz}$ 				& 217\,$\pm$\,37 				& $\mu$Jy\\
$r_{\rm FWHM}$				& 2.6$_{-2.6}^{+4.5}$ 			& arcsec\\
$L_{\rm 1.4~GHz}$ 				& 2.32\,$\pm$\,0.40 				& 10$^{25}$~$\mu^{-1}$~W~Hz$^{-1}$\\
$L_{\rm 8-1000~\mu m}$ 	& 5.6$^{+4.1}_{-2.4}$					& 10$^{13}$~$\mu^{-1}$~L$_{\odot}$\\
SFR 							& 9.4$^{+7.4}_{-4.9}$			& 10$^{3}$~$\mu^{-1}$~M$_{\odot}$~yr$^{-1}$	\\
$T_{d}$ 						& 24$^{+7}_{-5}$ 				& K\\
\hline
\end{tabular}
\label{table}
\end{center}
\end{table}

\section{The total infrared luminosity}
\label{ir}
The radio luminosity density that we measured from the WSRT observations can be used to determine the total IR luminosity of MM\,18423+5938 via the well established radio--FIR correlation \citep*{condon91}, which shows a tight relation at $z < $~0.5 (\citealt{price92,yun01,bell03,jarvis10}), with possibly a slight evolution out to redshift 4.4 \citep{garrett02,beswick08,ibar08,murphy09,seymour09,bourne10,ivison10a,ivison10b,kovacs10,michalowski10,sargent10a,sargent10b}. The radio--FIR correlation is usually defined by the parameter $q_{\rm IR}$, which relates the radio luminosity density and the total IR luminosity by,
\begin{equation}
q_{\rm IR} = \log_{10} \left( \frac{L_{\rm 8-1000~\mu m}}{[3.75\,\times\,10^{12}~{\rm Hz}]~L_{\rm 1.4~GHz}}   \right).
\end{equation}
\citet{ivison10b} found a typical value of $q_{\rm IR} =$~2.40\,$\pm$\,0.24 from a sample of 250~$\mu$m-selected galaxies ($S_{\rm 250~\mu m}\ga$~20~mJy; 5$\sigma$) that were observed with {\it Herschel}. This sample is comprised of 39 sub-mm galaxies at $z<$~3 with photometry in 12 bands between 24 and 1250~$\mu$m, and represents the largest sample of sub-mm galaxies with well sampled rest-frame IR spectral energy distributions studied to date. Using this definition of the radio--FIR correlation we find that the total IR luminosity of MM\,18423+5938 is $L_{\rm 8-1000~\mu m}=$~5.6$^{+4.1}_{-2.4}$\,$\times$\,10$^{13}$~$\mu^{-1}$~L$_{\odot}$.

This radio-derived total IR luminosity is $\sim$9 times smaller compared to what was found by \citet{lestrade10} from the IR spectral energy distribution of MM\,18423+5938. A possible explanation for this difference is the expected evolution in $q_{\rm IR}$ with redshift that results from an attenuation of the radio emission due to inverse Compton scattering off of cosmic microwave background photons \citep{condon92}. However, in contrast to this theoretical prediction, the observed value of the radio--FIR correlation has remained fairly constant with redshift (within the uncertainties). For example, the value of $q_{\rm IR}$ used here is consistent with that found for sub-mm galaxies at $z\sim$~4.4 ($q_{\rm IR} =$~2.16\,$\pm$\,0.28) by \citet{murphy09}, albeit with a much smaller sample size. The value of the radio--FIR correlation parameter that is implied from our radio observations and the \citet{lestrade10} total IR luminosity is $q_{\rm IR} =$~3.34\,$\pm$\,0.11, which is 4$\sigma$ from the results of \citet{ivison10b}.

Alternatively, the large difference in the derived total IR luminosities may arise from the method applied by \citet{lestrade10}, who used a Milky Way dust model with a large grain (cold) component at 45 K, a small grain (warm) component at higher temperature and a polycyclic aromatic hydrocarbon (PAH) component to describe the MM\,18423+5938 spectral energy distribution. The PAH component was constrained by an upper-limit at 24~$\mu$m from {\it Spitzer}, but since this component only dominates the rest-frame spectral energy distribution of the \citet{lestrade10} model below 8~$\mu$m, and is several orders of magnitude weaker than the other components, any change in the PAH strength will not significantly effect the total IR luminosity measured for MM\,18423+5938. The small dust grain component was constrained by the 70~$\mu$m detection with {\it Spitzer} and by the upper-limits at 60 and 100~$\mu$m from {\it IRAS}. However, as already discussed by \citet{lestrade10}, there is a 9~arcsec offset between the 1.2~mm position of MM\,18423+5938 and the possible 70~$\mu$m counterpart. Due to the higher astrometric precision of our radio data, we find this offset to be 10.8\,$\pm$\,1.9~arcsec and can now clearly establish that these two sources are not associated with each other at the 5.7$\sigma$ level (note that our analysis also includes a 1.7 arcsec uncertainty for the {\it Spitzer} 70~$\mu$m position). By using the limiting flux density of the 70~$\mu$m imaging as a constraint ($S_{\rm 70~\mu m}<$~5.7~mJy; 1$\sigma$), we find that the luminosity of the small dust grain component must be at least a factor of $\sim$5 smaller than was originally found by \citet{lestrade10}. However, this is not sufficient to fully explain the difference between the total IR luminosities derived using the different methods, even though the small dust grain component dominated the rest-frame spectral energy distribution between $\sim$8--30~$\mu$m and contributed $\sim$70 per cent to the total emission of the \citet{lestrade10} model.

\begin{figure}
\begin{center}
\setlength{\unitlength}{1cm}
\begin{picture}(6,7.6)
\put(-1.4,-0.5){\includegraphics{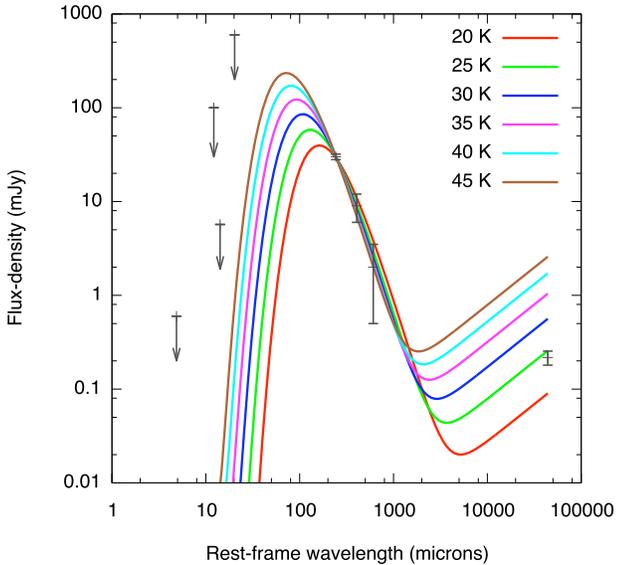}}
\end{picture}
\caption{A fit to the observed spectral energy distribution of MM\,18423+5938 using a single temperature modified blackbody spectrum with a dust emissivity of $\beta=$~1.5\,$\pm$\,0.5. The predicted radio emission for each temperature model is calculated from the radio--FIR correlation and assumes a power-law spectrum, $S_{\nu} \propto \nu^{-0.8}$. The data points are shown by the crosses. Given the large uncertainties introduced by the radio--FIR correlation and the unknown dust emissivity, we find that cold dust temperatures of $\sim$24$^{+7}_{-5}$~K are needed to reproduce the radio-derived total IR luminosity.} 
\label{sed}
\end{center}
\end{figure}

The large dust grain component provides most of the total IR luminosity from sub-mm galaxies and dominates the spectral energy distribution of MM\,18423+5938 from at least $\sim$30--1000~$\mu$m. To investigate the properties of this component, we have used a single temperature modified blackbody spectrum,
\begin{equation}
S_{\nu} \propto \frac{\nu^{3+\beta}}{{\rm exp}\,(h\nu/kT_d) - 1},
\end{equation}
where $T_d$ is the dust temperature of the large dust grain component and $\beta$ is the emissivity, which we assume is 1.5\,$\pm$\,0.5. We first attempted to fit a wide range of dust temperatures between 20 to 45~K and found that $T_d$ (and hence the luminosity) is not at all well constrained by the three mm detections of MM\,18423+5938 and the four non-detections at shorter wavelengths (the reduced $\chi^2$ of these fits were typically around 0.7 to 0.8). This is because the peak and the general shape of the spectral energy distribution, which are defined by the dust temperature, are not well sampled by the available photometric data (see Fig.~\ref{sed}).

We have attempted to place some constraint on the temperature of the large dust grain component by measuring the total IR luminosity for various dust temperature models and comparing with the result from the radio--FIR correlation. We calculated the FIR luminosity by integrating the best-fitting modified blackbody spectrum between 40 and 120~$\mu$m. A bolometric correction term was used to convert the FIR luminosity to the total IR luminosity, such that, $L_{8-1000~\mu m} = L_{40-120~\mu m}~\times$~1.91 \citep{dale01}. We find that the large dust grain temperature of MM\,18423+5938 has to be $T_d\sim$~24$^{+7}_{-5}$~K with a peak wavelength at around 150~$\mu$m to reproduce the radio-derived total IR luminosity. The error in the dust temperature includes the uncertainty introduced by the radio--FIR correlation and the unknown dust emissivity.  Recent results from {\it Herschel} have shown that high redshift sub-mm galaxies have cold dust temperatures of 34\,$\pm$\,5~K (\citealt{chapman10}; see also \citealt{magnelli10} who find 36\,$\pm$\,8~K), and peak rest-frame wavelengths of around 100~$\mu$m. The predicted cold dust temperature of MM\,18423+5938 is lower than these values, which suggests that the star-forming conditions of this galaxy are not as extreme as the typical sub-mm galaxy at redshift 2 (see below for further discussion). 

The single temperature modified blackbody approximation that we have used here is probably too simple to fully characterize the cold dust component of the spectral energy distribution. More sophisticated radiative transfer models (e.g. \citealt{dale02}) for the full spectral energy distribution of MM\,18423+5938 can be tested once additional data have been collected. However, our analysis has shown that with the available photometric data, the rest-frame spectral energy distribution between 8--1000~$\mu$m is not sufficiently well constrained to independently yield a reliable total IR luminosity. Observations at 100 to 850~$\mu$m with, for example {\it Herschel} and SCUBA2, will be needed to properly constrain the spectral energy distribution and the dust temperature of MM\,18423+5938 (e.g. \citealt{chapman10}) so that the total IR luminosity can be measured in this way.

Finally, as an independent check of our radio-derived total IR luminosity for MM\,18423+5938 we have used the relation between the CO (3$-$2) luminosity and the FIR luminosity, which for high redshift sub-mm galaxies has a ratio of $L_{\rm 40-120~\mu m} /  L'_{\rm CO~(3-2)} =$~194\,$\pm$\,20 L$_{\odot}$ / K~km\,s$^{-1}$~pc$^2$ \citep{iono09}. The CO (3$-$2) transition was not observed by \citet{lestrade10}, but based on their CO spectral energy distribution this transition is predicted to have a flux-density of $\sim$18~mJy. We find for our radio-derived FIR luminosity that the CO (3$-$2) transition should have a flux-density of $\sim$12$^{+8}_{-5}$~mJy, assuming the same 175 km\,s$^{-1}$ line width as for the CO (4$-$3) transition. This is in good agreement with what is predicted for the CO (3$-$2) transition from the spectral energy distribution. Therefore, we find that the best estimate of the total IR luminosity of MM\,18423+5938 currently comes from the radio imaging presented here, under the assumption that the radio--FIR correlation holds out to redshift 4.

\section{The properties of MM\,18423+5938 and the need for gravitational lensing}
\label{lensing}

The radio luminosity density of MM\,18423+5938 at 1.4 GHz is about 1.5 orders of magnitude higher than the typical near-infrared selected star-forming galaxy found between redshift 1.42 and 2 \citep{bourne10} and is about 9 times higher compared to the population of radio selected sub-mm galaxies found at $z\sim$~2.5 \citep{chapman05}. The radio luminosity density is also 3-8 times higher compared to 4 out of 5 (unlensed) spectroscopically confirmed sub-mm galaxies at $z\sim$~4 with radio detections; the exception being the sub-mm galaxy GN20 \citep{daddi09a}, which is known to host a radio-loud AGN. The unusually high radio luminosity density could therefore suggest that the emission from MM\,18423+5938 is predominately due to a radio-loud AGN. However, since the CO spectral energy distribution peaks at $J=$~5 and is well fit by a single low excitation temperature model, strong heating by a central AGN can be ruled out.

The radio-derived total IR luminosity of MM\,18423+5938 is 6-8 times larger than what was found for typical sub-mm galaxies at redshift $z\sim$~2.5 \citep{kovacs06,magnelli10,chapman10}, but is consistent with the extreme values ($L_{\rm 8-1000~\mu m}\sim$~3\,$\times$\,10$^{13}$~L$_{\odot}$) reported for hyper-luminous sub-mm galaxies out to $z\sim$~2 \citep{casey10}. However, these extreme luminosity galaxies also have cold dust temperatures of $T_{d}=$~52\,$\pm$\,6~K, which is significantly higher than what we found for MM\,18423+5938. Using the relation between the total IR luminosity and the star-formation rate by \citet{kennicutt98}, the radio-inferred star-formation rate of MM\,18423+5938 is found to be SFR ~= 9.4$^{+7.4}_{-4.9}$\,$\times$\,10$^{3}$~$\mu^{-1}$~M$_{\odot}$~yr$^{-1}$. This is for a continuous starburst model of 10 to 100 Myr duration, is valid over the mass range 0.1 to 100~M$_{\odot}$ and assumes a \citet{salpeter55} initial mass function. Again, the star-formation rate is much higher than what has been found for the sub-mm galaxy population at $z\sim$~2.5 (400~M$_{\odot}$~yr$^{-1}$; e.g. \citealt{coppin08}), and is not consistent with the moderate excitation temperature implied by the CO spectral energy distribution.

Gravitational lensing can be used  to explain these contradictions between the observed properties (i.e. the radio luminosity density, total IR luminosity and star-formation rate) and the relatively low cold dust and CO excitation temperatures. This is because the temperatures are related to the shape of the spectral energy distribution (either for the continuum or CO spectral line transitions) and are not effected by gravitational lensing, providing there is little or no differential magnification, whereas the luminosities scale with the magnification. Since the total IR luminosity and star-formation rate that have been calculated from the radio imaging are $\sim$9 times lower than those found by \citet{lestrade10}, the need for a very high gravitational lens magnification ($\sim$300) to explain the emission from MM\,18423+5938 is no longer required. By assuming a reasonable magnification of between 10 and 50 from a possible gravitational lens, the predicted unlensed total IR luminosity and star-formation rate of MM\,18423+5938 decreases to 1.1--5.6\,$\times$\,10$^{12}$~L$_{\odot}$ and 190--940~M$_{\odot}$~yr$^{-1}$, respectively. Similarly, the unlensed radio luminosity density would decrease to 0.5--2.3\,$\times$\,10$^{24}$~W~Hz$^{-1}$. The range of assumed lensing magnifications is consistent with what has been found for other star-forming galaxies at high redshift that are gravitationally lensed (e.g. \citealt{negrello10}). 

High sensitivity observations at sub-arcsec resolution with, for example, the Expanded Very Large Array (EVLA) or the Multi-Element Radio Linked Interferometer Network ({\it e}-MERLIN) could be carried out to image the radio emission from MM\,18423+5938 and to confirm the lensing hypothesis for this system. Such observations would allow the actual magnification from the gravitational lens to be determined and find out if there is any contaminating radio emission from the foreground gravitational lensing galaxy (e.g. \citealt{berciano-alba10,volino10}). Further observations at milliarcsecond resolutions with very long baseline interferometry could be used to determine whether there is any contribution to the total radio flux-density from an AGN (e.g. \citealt{garrett01,biggs10}).

\section{summary}

MM\,18423+5938 at redshift 3.9296 was thought to have an extremely large total IR luminosity (4.8\,$\times$\,10$^{14}$~$\mu^{-1}$~L$_{\odot}$; \citealt{lestrade10}), and based on this was believed to be highly magnified by a gravitational lens. We have presented new imaging at 1.4 GHz that detects the galaxy at radio wavelengths for the first time and determines a more precise position. Using the well-established radio--FIR correlation, we find that the total IR luminosity of this galaxy is $\sim$9 times lower than what was previously calculated from an analysis of the IR spectral energy distribution, which is considered unreliable due to the limited photometric data that is currently available. In particular, we find that the object detected at 70~$\mu$m, which partly led to the high total IR luminosity, is not associated with MM\,18423+5938. The radio-derived properties are still quite large ($L_{8-1000~\mu m}$~=~5.6$^{+4.1}_{-2.4}$\,$\times$\,10$^{13}$~$\mu^{-1}$~L$_{\odot}$ and SFR~=~9.4$^{+7.4}_{-4.9}$\,$\times$\,10$^{3}$~$\mu^{-1}$~M$_{\odot}$~yr$^{-1}$) given the moderate excitation temperature of the CO lines and the low cold dust temperature ($T_d\sim$~24$^{+7}_{-5}$~K). Therefore, we conclude that MM\,18423+5938 is probably gravitationally lensed, as \citet{lestrade10} suggested. We find that extreme magnification factors are no longer required; a reasonable lensing magnification between 10--50 would give a star-formation rate and total IR luminosity that are consistent with a typical sub-mm galaxy at redshift $\sim$2.

\section*{Acknowledgments}
The Westerbork Synthesis Radio Telescope is operated by the ASTRON (Netherlands Institute for Radio Astronomy) with support from the Netherlands Foundation for Scientific Research (NWO). We would like to thank Gyula Jozsa for the prompt scheduling of the observations and for preparing the observing files, Ger de Bruyn for advice on analysing the data, and Scott Chapman, Daisuke Iono and Stephen Serjeant for useful discussions. FV and OW are supported by the Emmy-Noether-Programme of the Deutsche Forschungsgemeinschaft, reference WU 588/1-1.

\bsp

\label{lastpage}

\end{document}